\documentclass[sigconf, screen=true, authordraft=False]{acmart}

\usepackage{tabularx}
\usepackage{dcolumn}
\newcolumntype{d}[1]{D{.}{.}{#1}}
\usepackage[caption=false]{subfig}

\AtBeginDocument{
  \providecommand\BibTeX{{
    \normalfont B\kern-0.5em{\scshape i\kern-0.25em b}\kern-0.8em\TeX}}}
\setcopyright{none}
\acmYear{}
\acmDOI{}
\acmConference[cs.HC]{cs.HC arXiv.org}{2023}{arXiv.org}
\acmISBN{}

\begin{document}

\title{Composing Neural Networks with Direct Manipulation}
\title{The Unadopted Benefits of Graphical Programming for Data-Driven Software}
\title{Usability and Adoption of Graphical Data-Driven Development Tools}

\author{Thomas Weber}
\orcid{0000-0002-6894-605X}
\affiliation{
 \institution{LMU Munich}
 \streetaddress{Frauenlobstr. 7a}
 \city{Munich}
 \country{Germany}
 \postcode{80337}
}
\email{thomas.weber@ifi.lmu.de}

\author{Sven Mayer}
\orcid{0000-0001-5462-8782}
\affiliation{
 \institution{LMU Munich}
 \streetaddress{Frauenlobstr. 7a}
 \city{Munich}
 \country{Germany}
 \postcode{80337}
}
\email{sven.mayer@ifi.lmu.de}

\renewcommand{\shortauthors}{Weber and Mayer}

\begin{abstract}
Software development of modern, data-driven applications still relies on tools that use interaction paradigms that have remained mostly unchanged for decades. While rich forms of interactions exist as an alternative to textual command input, they find little adoption in professional software creation. In this work, we compare graphical programming using direct manipulation to the traditional, textual way of creating data-driven applications to determine the benefits and drawbacks of each. In a between-subjects user study (N=18), we compared developing a machine learning architecture with a graphical editor to traditional code-based development. While qualitative and quantitative measures show general benefits of graphical direct manipulation, the user's subjective perception does not always match this. Participants were aware of the possible benefits of such tools but were still biased in their perception. Our findings highlight that alternative software creation tools cannot just rely on good usability but must emphasize the demands of their specific target group, e.g. user control and flexibility, if they want long-term benefits and adoption.
\end{abstract}



\begin{teaserfigure}
    \centering
    \includegraphics[width=\linewidth]{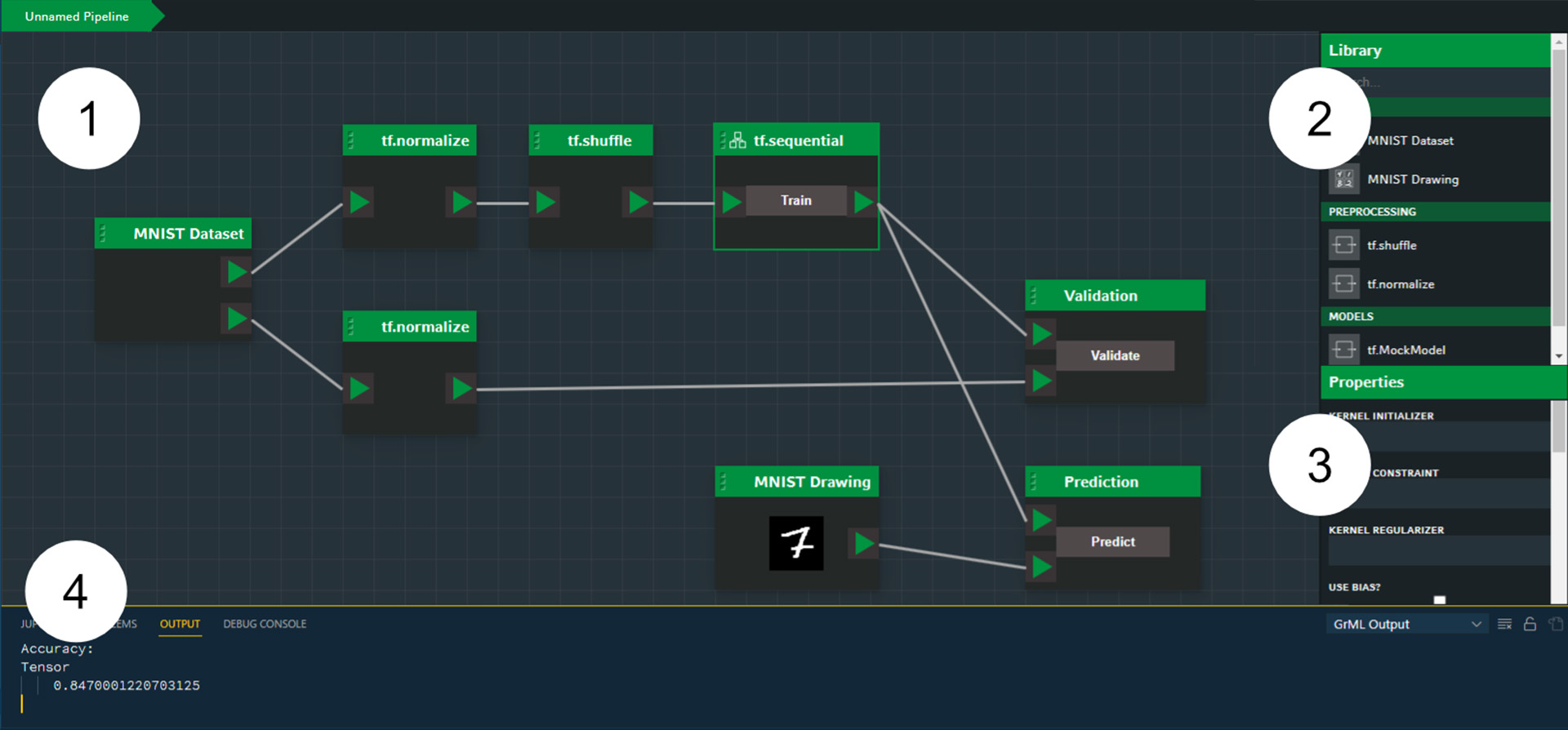}
    \caption{The graphical programming interface we created for this study to represent rich interaction with direct manipulation. It consists of a primary canvas (1) on which blocks can be placed. These blocks are dragged from the block library (2) and can have additional properties which can be edited (3). Any output is displayed in a dedicated area at the bottom of the screen (4).}
    \label{fig:grml-screenshot}
\end{teaserfigure}

\maketitle

\section{Introduction}

Machine Learning, particularly using Neural Networks, has enabled many improvements in fields  which are critical to everyday life in our society, like healthcare~\cite{Cohen2019MEDICAL}, communication~\cite{COMMUNICATION}, and security~\cite{SECURITY1,SECURITY2}. Therefore, neural networks are an essential tool for software developers going forward.

Yet, while these systems become increasingly complex and hard to understand, an issue exacerbated by their opacity~\cite{NEW42,HOFFMAN2018}, how developers create them is not much different from how other traditional software has been created for decades: the probably most popular tool, Jupyter notebooks\footnote{\url{https://jupyter.org}}, and many more are fundamentally still textual command input. Meanwhile, there are many alternative forms of interaction for programming, which could be applied here. Graphical programming in particular is an interesting candidate, given the nature of Machine Learning systems where data visualization often plays a large role, but many other systems also rely on some form of visual presentation for different goals like interpretability~\cite{ZHANG2018} and explainability~\cite{LIME,ALICIOGLU2022}, optimization~\cite{HYPERTUNER,HYPERTENDRIL}, debugging~\cite{AFYAL2021,alsallakh2021debugging} and many more. Graphical Programming systems like RapidMiner~\cite{RAPIDMINER1}, KNIME~\cite{KNIME1}, or Orange~\cite{ORANGE} fully embrace this and omit textual programming for the most part. And yet, most developers still fall back to their text editors or Jupyter notebooks, which raises the question whether these alternative systems are inferior or more importantly what they are missing.

In this paper, we investigated the behavior of programmers in these contexts to determine how and which aspects of graphical programming are beneficial and when developers prefer textual programming. To this end, we performed a between-groups user study (N=18), where participants had to create a Machine Learning system for image classification either traditionally with code or using a simple graphical programming tool, which we designed and developed specifically for this investigation. This tool allows for the graphical composition of Neural Networks via direct manipulation of its layers. We set this tool in contrast to the de facto standard tool, Jupyter notebooks, in which participants also designed new neural networks. This allowed us to understand the differences from a qualitative and quantitative perspective. Thus, addressing the question of what alternative systems are missing today and how we can improve them.

Our results show that graphical programming -- for our task -- is, in fact, already faster and produces less workload.
While it is therefore viewed positively, participants still expect that they work better with the traditional, text-based editor. This feedback provides insights into the potential benefits of graphical programming for data-driven development, but also highlights a disparity in users' subjective perceptions. Giving developers a high degree of control and flexibility over their tools appears to be necessary to ensure the success of graphical development tools. Based on our results, we argue that building a hybrid between text editors and graphical programming may be a way to consolidate the different benefits and expectations in the future.

\section{Related Work}
Richer interactions than text input are by no means new for software development in general or data-driven development in particular.
Direct manipulation as one of such rich interactions was, in fact, originally proposed by \citet{Shneiderman81} as an alternative to programming languages and command-based control of computers. Since then, research and industry have come up with many different forms of interaction.
Yet, many of the ideas specific to professional software engineering and, more recently, data-driven development have only found limited adoption in practice.
Anecdotal evidence suggests that particularly more radical changes that fundamentally change how developers interact with the software they create, fare poorly. One such drastically different interaction paradigm is graphical programming, which replaces text-based commands with visual representations.

While not embraced by the broad majority, graphical programming still succeeded in some niche application domains for professional software engineering. Often these are areas where the software is and needs to fulfill very strict requirements in timing, safety, etc.~\cite{EMBEDDED, PERFORMANCE, PERFORMANCE2, VERIFICATION}.
So from the outside perspective, the benefits of graphical programming, particularly for complex and critical systems, are apparent. However, given the adoption rate, from the inside perspective, the point of view of developers, these benefits seemingly are not yet convincing.

An early article by \citet{PETRE95} already investigated this disparity and compared the benefit of graphical programming for novices and experts where she describes some of the benefits of that arise from the rich interaction like accessibility, memorability, etc. However, the article also emphasizes the challenges which arise from the different needs of novices and experts. While graphical programming is often advocated for beginners, \citet{PETRE95} instead describes how they would rather benefit from more constrained interactions. The ability to grasp and utilize graphical representation is something that also requires some expertise and poorly designed visualizations can easily be confusing or overwhelming.

\citet{ERWIG1995HVLP} also provide an early discussion of the benefits of visual languages and their integration with the more common textual counterpart in heterogeneous visual languages for general programming activities.

\citet{MDEINDUSTRYSTUDY} later reviewed the state in the early 2000s in the context of Model-driven development tools and practices, which are often graphical in nature. At that point they found adoption to be limited and only at the small scale and evaluations to be lacking. Since then, many projects in which new graphical programming tools have been created now also conduct user studies to evaluate them, especially the application for children and in education~\cite{CHENWANG2011,WANG2013,HORN2009,SAPOUNIDIS201967} and end user programming~\cite{EUDSURVEY,EUDEXAMPLE} but also beyond.

After all these years, graphical programming remains an active research interest, as seen in the literature review by \citet{SELWYNSMITH2022} and recent examples like  the work of \citet{HOMER2023}, who investigates the use of graphical programming for data-flow programming.

\subsection{Graphical Programming for Data-Driven Applications}
Data-driven applications have a high degree of inherent complexity, so graphical programming could be a viable route to mitigate this. Consequently, academia and industry have come up with some systems that try to leverage this interaction to simplify development.

Tools like RapidMiner exist for two decades now~\cite{YALE} and have been applied in several domains and case studies (e.g., \cite{RAPIDMINER_EXAMPLE1,RAPIDMINER_EXAMPLE2,RAPIDMINER_EXAMPLE3,RAPIDMINER_EXAMPLE4}).
RapidMiner and other tools (e.g., \cite{ORANGE, KNIME1}) frequently visualize the data-flow in data-driven applications as a sequence of blocks, which represent different steps in the processing.
With this representation, they are able to make a relatively abstract process easier to grasp, while hiding unnecessary details. Of course, many of these details are necessary for experienced users, so there are different interaction patterns for accessing them, from popups to the common ``bento box'' \cite{ZOOMABLE} layout to hierarchical structures.
They also often provide additional visualizations as support, e.g., for data exploration or to display output metrics.
While not specifically targeted at novices or experts, they are still often considered more suitable for novices~\cite{DEPTHINSIGHTFORDATASCIENTIST} or non-programmers~\cite{MACHINELEARNINGFORNONPROGRAMMERS}.

\citet{weber22muc} provide a further overview of tools for data-driven software development from the last decade. Based on their literature review, they also draw the conclusion that graphical programming is one area that could improve data-driven development, but has not yet managed to gain enough traction and practical use.

Instead, for developers, Jupyter notebooks so far seem the tool of choice~\cite{RANDLES2017}. From an interaction design point of view, these notebooks are still different from the setup common in traditional software development with either a text editor and tools for compilation and execution, or the Integrated Development Environment (IDE) which combines them. Jupyter notebooks rather focus on small code fragments and short execution and feedback cycles, very similar to a REPL (Read-Evaluate-Print-Loop) or interactive shell environment.
In this sense, Jupyter notebooks can be seen as a regression towards the interaction as one would have with a terminal with superior back-tracing and presentation. Still, it is the development tool of choice particularly in the domain of data-driven applications, so any novel and improved tools will measure up to it.

Furthermore, the Jupyter environment provides infrastructure for extending it, which not only contributed to its success across many  programming languages, but has also enabled researchers to extend the fairly conservative forms of interaction.
\citet{THEFUTUREOFNOTEBOOK} and \citet{zhao2022oden} both attempted to enhance the Jupyter programming environment with visualizations that allow developers to switch between the traditional and graphical representations. In the user studies, which they conducted, they found that professional users enjoy the flexibility of switching between different representations, but they also received critical feedback indicating that experts struggle with some interactions. In the ODEN tool by \citet{zhao2022oden}, for example, the participants were sometimes confused by the mapping between graphical and textual representation, which prompted them to implement a ``calm mode''. Clearly, designing tools that are suitable for developers of data-driven applications is not a trivial task. To ensure that they can provide the benefit they promise, we must first better understand what interactions are suitable and what users prefer in a given context.

\section{A Graphical Programming Editor for Neural Networks}
As existing graphical programming tools have an overwhelming amount of features, we decided to work with a prototypical implementation of a graphical programming tool. Therefore, we limit the volume of features and interactions to a minimal, comparable subset and completely control what kind of interactions were possible during our study. A custom implementation also allowed us to instrument the tool, giving us much more detailed information about the interactions. Nonetheless, the tool is inspired by and based on established tools like RapidMiner~\cite{RAPIDMINER1} or KNIME~\cite{KNIME1} and their block-based interaction: instead of instruction as lines of code, the functions for processing the data are displayed as blocks. These blocks can be arranged in space via drag and drop. These blocks can then be combined into a sequence of operations. This is done via direct manipulation by dragging a connection from one block to another. Combining blocks like this lends itself to the domain of data-driven applications, since a mostly sequential pipeline is a typical structure.

This is the primary canvas for creating programs (see \autoref{fig:grml-screenshot}).
In addition, similar to how Jupyter notebooks display their output below each cell, the UI had a panel for displaying any program output in a consistent location.
Besides this, the tool offers a library of blocks, which we grouped into categories like \textit{Datasets}, and \textit{Models}. Furthermore, each block can have parameters, just like a function call has parameters. These are displayed in a third panel as input fields depending on the parameter's type, e.g., checkboxes for boolean parameters.

Since we base the task on a tutorial that uses the Keras layers API, we implemented a subset of its functionality as blocks in our application. This was convenient as the TensorFlowJS library, which we used internally for the UI, uses mostly the same API, which allowed us to use the same instructions and functions in both experimental conditions. The blocks we implemented included all the necessary models, layers, and data processing steps for the tutorial and a few additional distractors. Each of these blocks used the same parameters as listed in the Keras/TensorFlowJS documentation. Furthermore, we added the MNIST dataset as a block, similar to how it is available in the \textit{tensorflow\_datasets} Python package used in our conventional tutorial, and two blocks to enable interactive prediction of new inputs.

In the Keras API, a typical model is composed of multiple layers, i.e., a function composed of sub-functions. For this reason, our tool offers the possibility of composite blocks, i.e., blocks that consist of a sequence of subordinate blocks. Double-clicking such a composite block opens up a new editor view for arranging its components. Breadcrumb navigation at the top of the main canvas allows for navigation along this hierarchy. In our scenario, we used this type of block to create a \textit{sequential} model from different types of layers, like \textit{dense} or \textit{flatten} layers.

We implemented this UI as an extension for Microsoft's Visual Studio Code editor in TypeScript.
To complete the task we provided, the participants only interacted with our extension and no further aspects of Visual Studio Code. The source code for the extensions is part of the supplementary material and is ready to be part of a public code repository.

\section{User Study}
To understand how different types of tools and interactions paradigms affect the programmers, we conducted a between-subject study. In this study, participants had to create a simple machine learning system using a neural network for classifying the MNIST handwritten digits \cite{lecun1998mnist} based on a set of instructions and provide feedback on their experience and preferences using a questionnaire. Additionally, we conducted semi-structured interviews with the participants to get additional in-depth feedback.

In the study, one group of participants created their neural network using the established interaction of textual command input, using \textit{Jupyter notebooks}, the de-facto standard tool for data science programming. The second group created the same system using an equivalent set of instructions, but with the graphical programming tool described above. With this, we were able to focus our study and the interface to those interactions in the intersection of these two different tools.

\subsection{Task}
The instructions were designed to follow the general outline of a typical machine-learning tutorial. For this reason, we took an existing example from the TensorFlow documentation\footnote{\url{https://www.tensorflow.org/datasets/keras_example}} and adapted it, where necessary, for our study.

The most notable change is that we split the training and validation, which in the original are a single function call, into two separate steps. The tutorial version for the graphical programming group received some additional changes: we replaced all code examples with screenshots of our tool in the respective state, and we had to rephrase some sentences such that they reflect the difference in interaction, which, for example, meant replacing occurrences of \textit{to type} or \textit{to write} with more suitable verbs.

We also opted to keep the visual presentation of the tutorial as close to the original as possible, but removed any clutter and unnecessary links from the site to ensure that participants were focused on the core instructions. Any external links were also removed to keep participants on the tutorial page.

\subsection{Procedure}
After the participants arrived, we gave them a small introduction and asked for their consent to record the data. The study starts with a brief introduction of the tool which participants get to use. For \textit{Jupyter Notebooks}, this includes information on how to write and execute code and how to extend the notebook with new cells. For \textit{Graphical Programming}, we introduce the panels described above and the direct manipulation via drag and drop.

After this, participants received access to the instructions, and we asked them to work through these at their own pace. The whole task was scheduled to take about 15 minutes. As per the study invitation, participants were aware that we scheduled the duration of the study at 45 minutes and the programming task was expected to take roughly 15 minutes, but we did not rush them to finish the programming task or cut them off after the allotted time.

At the end of the programming instructions, we included a link to the survey, which constituted the last part of the study for most participants. This survey included both qualitative and quantitative questions, as we describe below. A subset of five participants was able to accommodate time for an additional semi-structured interview  after the survey, in which we collected additional qualitative feedback.

\subsection{Survey and Interview Guidelines}

The survey was separated into three core parts: the first part was concerned with a consent form and information about data protection, followed by questions about demographic data and background knowledge and expertise. We used the scale from \citet{EDISON2003} to record a general attitude towards technology. In addition, we queried participants about their experience with the technologies specific to our study, i.e., Jupyter notebooks, graphical programming, and different topics about machine learning.

The second part of the survey asked for feedback on participants' programming tasks and tools. First, we used quasi-continuous, i.e., fine-grained, Likert scales with an agree-disagree scale \cite{chimi2009likert} for a series of general statements about the task and the tool.
Second, we asked participants to rate the tool based on several categories based on Nielsen's usability heuristics \cite{nielsen2005ten} where applicable and some additional categories (see \autoref{fig:heuristics}). Finally, we used the Systems Usability Scale \cite{brooke1996sus} and the raw NASA Task Load Index \cite{hart2006nasa} as further metrics.
Participants could also provide additional qualitative feedback using free text fields regarding the tool's benefits, limitations, and drawbacks.

After providing this feedback, we showed the participants a screen recording of their task being executed in the alternative tool, Jupyter, for participants who used the graphical tool and vice versa. Having watched this, we asked participants to compare what they had seen to what they had used, using the same categories as before. Here, too, participants could provide additional qualitative feedback on what they thought to be better or worse in this alternative tool.

As mentioned, we discussed their experience with several participants after they had completed the survey in a semi-structured interview. The interview questions first focused on the general experience during the study and what aspects of the task were particularly challenging. We then asked about the tool, its benefits and drawbacks, and under what circumstance or for which target group it might be most suitable.
Finally, we asked for feedback on how to make the machine learning programming easier to use in the everyday work context of the participants.

\subsection{Apparatus}
Participants completed both the survey and the programming task on a computer provided by us using a mouse and keyboard. This way, we ensured that the training phase of the machine learning system was using the same hardware, making it more consistent. The page with the instructions was displayed on a second screen to minimize the need for switching windows. Aside from the survey, additional information like the timing of interactions and the task result was automatically recorded in the background.

\subsection{Participants}
For this study, we recruited participants with prior experience with programming, particularly of data-driven applications. Knowledge of specific Machine Learning libraries or methods was not a prerequisite. We recruited them via a combination of personal and professional contacts and multiple mailing lists that specifically included people with programming backgrounds.

In total, 20 participants completed our study (4 female, 16 male). With a mean age of 29 ($SD = 5.2$) and six years of programming on average ($SD = 3.8$). Half of the participants had a background in computer science, while the other half came from STEM areas where data processing and the use of machine learning methods is common, for example, physics or biology research.
The participants were employed full-time in their positions, typically with a completed Master's degree (14), or were currently still pursuing a degree in these domains but worked part-time in projects involving data-driven applications (6). Based on their feedback, the participants had a generally positive attitude towards technology in general (average score: 76 of 100, $SD = 11$) and were at least somewhat familiar with all the specific technologies of our study, like machine learning or Jupyter notebooks.
While they knew of the MNIST data set, none of them reported to have completed the specific tutorial upon which our task was based. Four of the participants answered that they use graphical programming tools in their work, although not for creating data-driven applications.
We scheduled for a session of 45 minutes per participant for introduction, performing the task as well as completing the questionnaire, and, if time allowed it, additional feedback. Regardless of whether they required the whole time, participants were compensated with an equivalent of 10 US\$ for their contribution.

\begin{figure*}[t]
  \centering
  \includegraphics[width=\linewidth]{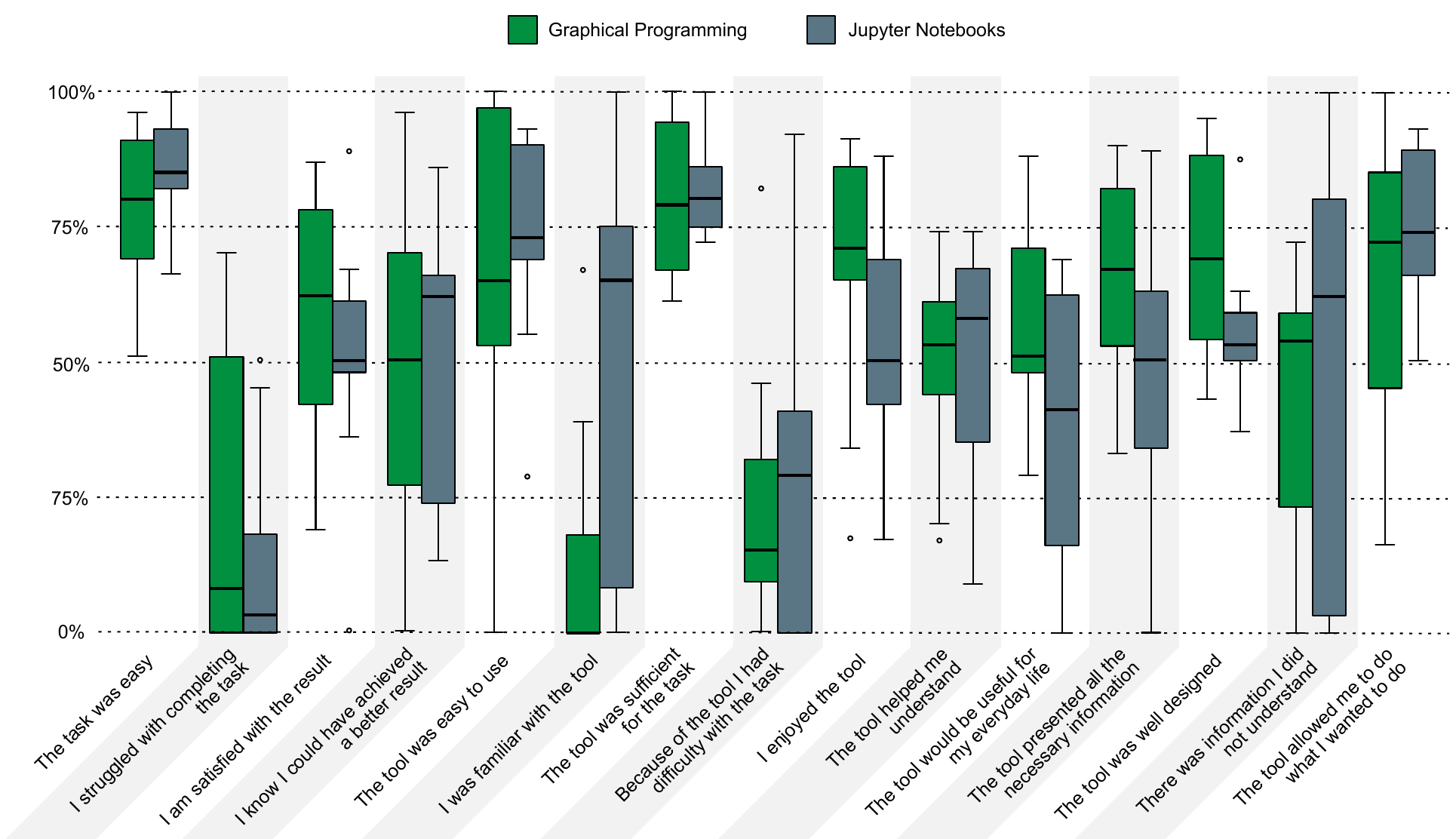}
  \caption{While there are no significant differences between the responses to various statements about the task and the tools, they show that the task was perceived very similarly.}
  \label{fig:statements}
\end{figure*}

\section{Results}

Of the 20 people who participated in our study, we excluded two from the data set, one from each group, after the first data screening, as their responses strongly suggested that they misunderstood parts of the instructions. The remaining data yielded the following insights.

As \autoref{fig:statements} shows, the task was perceived as easy to complete in both experimental conditions. Given the overall very similar feedback about the task, both in the survey and the interviews, we conclude that the task design was suitable, as it was not notably easier or better supported by any of the two tools.

The scales asking about the tools show some differences: while the graphical programming tool received slightly better feedback on enjoyment, how it supports understanding, and its overall design, these differences are not significant.

\subsection{Usability}
\begin{figure*}[t]
  \centering
  \vspace{-0.3cm}
  \includegraphics[width=\linewidth]{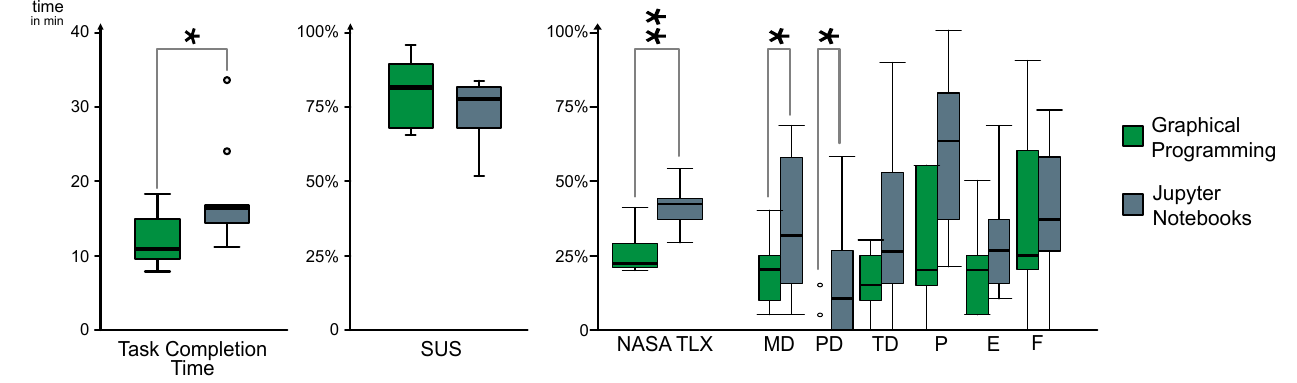}
  \vspace{-0.3cm}
  \caption{Task completion time and NASA TLX results show significant differences, while the SUS shows positive results for both UIs. (*:~$p < 0.05$, **:~$p < 0.01$)}
  \label{fig:quantitative}
  \vspace{-0.3cm}
\end{figure*}

In terms of usability Both tools scored well on the Systems Usability Scale \cite{brooke1996sus}, 81 of 100 ($SD= 11.7$) for the graphical tool and 74 ($SD = 10.4$) for Jupyter notebooks, see \autoref{fig:quantitative}. This difference is not significant.

The workload as measured by the raw NASA TLX \cite{hart2006nasa}, meanwhile, is significantly lower for the graphical tool (Shapiro-Wilk-test: $W = 0.934$, $p = 0.287$,  t-test: $t(17) = -4.027$, $p < 0.001$). Moreover, each of the six subscales of the NASA TLX is lower for the graphical tool individually, highlighting the reduction in workload even further (see \autoref{fig:quantitative}).

\begin{figure*}[t]
  \centering
  \includegraphics[width=\linewidth]{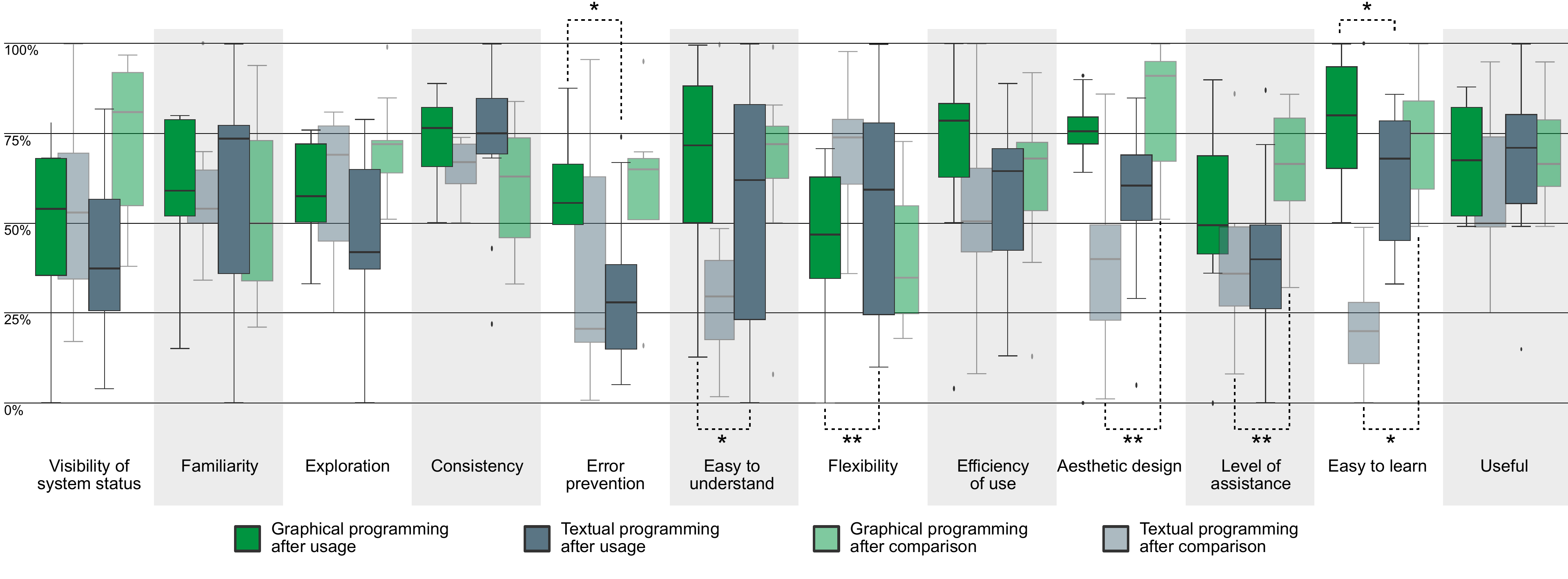}
  \caption{Each participant rated the tool they used and the alternative tool in comparison using the same properties and heuristics about usability (*: $p < 0.05$, **: $p < 0.01$)}
  \label{fig:heuristics}
\end{figure*}

Furthermore, the graphical tool was also rated slightly superior in appearance, ease of understanding and learning in the questions (see \autoref{fig:heuristics}). The participants also rated the graphical programming UI to be significantly superior in terms of \textit{error prevention} ($p = 0.023$) and how \textit{easy to learn} they considered it ($p = 0.025$), see \autoref{tab:stats} for a complete list).

\begin{table*}
    \centering
    \captionof{table}{Results of the statistical analysis of how participants judged the systems concerning usability heuristics (Single Judgment) and comparing the system they used to the alternative as seen in a screen recording (Comparative Judgment).}
    \small
    \label{tab:stats}
    \begin{tabularx}{\linewidth}{X*{5}{d{3}}l*{5}{d{3}}l}
        \toprule
         & \multicolumn{6}{c}{\textbf{Rating after usage}} &  \multicolumn{6}{c}{\textbf{Comparative Rating}}\\
         \cmidrule(r){2-7}\cmidrule(l){8-13}
         &  \multicolumn{2}{c}{\textbf{Normality test}} & \multicolumn{4}{c}{\textbf{Wilcoxon/t-test}} &  \multicolumn{2}{c}{\textbf{Normality test}} & \multicolumn{4}{c}{\textbf{Wilcoxon/t-test}} \\
         \cmidrule(r){2-3}\cmidrule(rl){4-7}\cmidrule(rl){8-9}\cmidrule(l){10-13}
        & \multicolumn{1}{c}{\textbf{W}}     & \multicolumn{1}{c}{\textbf{p}}     & \multicolumn{1}{c}{\textbf{W/t}}    & \multicolumn{1}{c}{\textbf{df}}     & \multicolumn{1}{c}{\textbf{p}}     & & \multicolumn{1}{c}{\textbf{W}}     & \multicolumn{1}{c}{\textbf{p}}     & \multicolumn{1}{c}{\textbf{W/t}}    & \multicolumn{1}{c}{\textbf{df}}     & \multicolumn{1}{c}{\textbf{p}}     &    \\
        \midrule
         Visibility of System Status           & 0.973 & 0.852 &  0.8   & 15.991 & 0.435 &   & 0.945 & 0.442 & -1.903 &  9.724 & 0.087 &    \\
         Familiarity        & 0.857 & 0.011 & 45.5   &        & 0.691 &   & 0.956 & 0.627 &  0.157 & 12.763 & 0.878 &    \\
         Exploration        & 0.910 & 0.085 &  1.283 & 13.907 & 0.221 &   & 0.943 & 0.428 & -0.858 & 11.273 & 0.409 &    \\
         Consistency        & 0.952 & 0.460 & -0.018 & 12.85  & 0.986 &   & 0.950 & 0.634 &  0.157 &  9.644 & 0.879 &    \\
         Error Prevention   & 0.953 & 0.480 &  2.480 & 15.638 & 0.025 & * & 0.915 & 0.141 & -1.636 & 12.345 & 0.127 &    \\
         Easy to Understand & 0.907 & 0.077 &  1.331 & 14.827 & 0.203 &   & 0.955 & 0.536 & -2.089 & 14.414 & 0.019 & *  \\
         Flexibility        & 0.973 & 0.848 & -0.474 & 13.387 & 0.643 &   & 0.920 & 0.168 &  3.212 & 13.915 & 0.006 & ** \\
         Efficiency of Use  & 0.900 & 0.050 &  1.133 & 15.852 & 0.274 &   & 0.964 & 0.740 & -0.556 & 10.509 & 0.590 &    \\
         Aesthetic Design    & 0.823 & 0.003 &   57.5 &        & 0.145 &   & 0.935 & 0.287 & -3.305 & 11.829 & 0.006 & ** \\
         Assistance         & 0.946 & 0.369 &  1.034 & 15.984 & 0.317 &   & 0.946 & 0.462 & -3.318 & 11.604 & 0.006 & ** \\
         Easy to Learn      & 0.951 & 0.439 &  2.517 & 15.279 & 0.023 & * & 0.927 & 0.191 & -2.384 & 14.7   & 0.031 & *  \\
         Useful             & 0.948 & 0.397 &  0.564 & 13.443 & 0.582 &   & 0.956 & 0.557 & -0.665 & 12.401 & 0.518 &    \\
         \bottomrule
    \end{tabularx}
\end{table*}

While these responses showed only moderate differences, we observe a stronger change when participants were asked to compare the interface which they used to the one they saw in the screen recording. After seeing the alternative tool, participants rated the graphical tool significantly more favorable concerning appearance, how it offered assistance, and how easy it might be to learn. The Jupyter notebooks, on the other hand, were rated positively for their perceived increased flexibility compared to the graphical tool. See \autoref{fig:quantitative}, \autoref{tab:stats} for additional details. \autoref{fig:quantitative} also shows the difference in opinions for participants for these categories, which further highlights how the comparison skews the perception of the participants.

\subsection{Task Completion Time}
We then compared the recorded task completion time for all participants. As \autoref{fig:quantitative} shows, completing the task using textual input on average takes longer by a significant amount (Mann-Whitney-U test, $p = 0.040$). While this is in part due to two outliers, the questionnaire responses from these two do not suggest that they considered the task more challenging or struggled more than the other participants.

\subsection{Qualitative Results}

Beyond these scales, we also queried participants for qualitative feedback on what aspects of the tools they saw as positive or as an impediment for adopting and using them. After an initial screening of the responses to the survey and the interview, we clustered them into five categories: responses specific to the task, the tool's appearance, understanding of the system, feature requests, and feedback on the interaction paradigms.

All nine participants who commented on the task described it as ``simple and easy (as it was) split into meaningful parts'' (P3). This is the case for both groups.
The feedback on the appearance of the graphical tool was positive, with multiple participants enjoying the visual presentation of the data flow via connections.
The focus on the blocks led one participant to note that the output can be ``easy to overlook'' (P15), though. The presentation of output close to the code as done in Jupyter, on the other hand, was considered to be convenient but constituted the only comment on the appearance of Jupyter notebooks.

Issues with understanding what was going on were related to general machine learning concepts. One participant expressed the opinion that ``(Jupyter) notebooks require prior knowledge'' (P6), whereas the graphical UI was described as especially suitable for beginners. To enable growth and learning, one participant suggested that ``a person learning could switch between'' (P6) graphical and text-based UI to use the presentation that suits them best and at any point in time. This flexibility was also a point of multiple participants asking for more information to be displayed on demand in the graphical UI, particularly for the parameters, as it was not always clear ``what parameters (...) were doing'' (P9). In addition to documentation, participants also favored more and particularly continuous feedback. Similar to how in Jupyter notebooks, each cell has its output adjacent, our participants wanted to see information about the success and output of blocks close to them in the graphical UI. In the Jupyter notebooks, the only indicator for a complete computation without output is a change in cell numbering -- an indicator that, according to participants, is easily overlooked, as we also observed during the study. Still, one participant praised the blocks, into which code is split in a notebook, as they help to structure longer code. This can lead to other issues, though, as ``boxes can be compiled in any order, meaning that code can crash if compiled in the wrong order,'' so there was a wish for information about data flow and notifications about changes with dependencies. If something does indeed not work, Jupyter shows the corresponding error messages. These messages were criticized by four participants to be ``cryptic'' (P1) and not helpful. However, this is not per se an issue of Jupyter, but of the Python programming language. The error message of the graphical UI, which use natural language, were praised by one participant to be good for ``understanding the way things worked'' (P15).

The key difference between the two systems lies in the form of interaction, though. Thus, participants provided a number of comments on the interaction after both using the system and seeing the alternative system.
Four participants mentioned the issue of typing errors, which can be hard to notice and typically lead to immediate errors in the code. Writing code, however, forces it to adhere to a strict sequence of instructions, which matches how it is executed.
On the other hand, the graphical presentation is not constrained like that, so more emphasis is put on the connections to indicate the sequence of execution. However, two participants highlighted that they liked the presentation with lines as connections. This also enables the tool to display data flow across steps that are not directly adjacent. In our task, an example of this is the validation data set. It is used fairly late in the process, which in code means that there are multiple code cells between its declaration and usage, whereas the blocks have a direct connection.

The interaction with drag and drop and direct manipulation to create the connections and the blocks were commented upon positively by seven of the participants. Grabbing the blocks from a library of predefined blocks was also mentioned as a desirable feature, as it reduces the need to recall API functionality from memory. While this does apply to the properties of selected blocks, two participants voiced confusion, as they were overwhelmed by the number of available parameters, even though they only needed to change one or two for the task.

\subsection{General Feedback}

Based on the feedback from the questionnaire and the interviews, we also collected some additional general feedback and insights from participants how these tools could become more useful in their work context to create data-driven systems.

\subsubsection{For Jupyter Notebook}
As described above, the primary area of criticism for the Jupyter notebooks were its error message. Participants not only wanted easier-to-understand text explanations, but particularly requested more precise feedback on the error location. Since cells can be executed independently of each other, errors in later cells can originate in prior ones. Tracking them back to their true origin was described as challenging in the current system, particularly when the number of cells and their size grows. In general, participants voiced interest in more detailed information about data flow in the notebooks.

Integration was another concern, as some participants saw Jupyter notebooks more as a tool for experimentation. When the software becomes very large, keeping it in one notebook becomes unfeasible. Integrating the functionality from one notebook with other, potentially legacy code, was considered challenging and needed a simple mechanism. While this can also apply to graphical programming tools, participants did not mention this concern in this context.

Inline access to documentation and auto-completion were also mentioned as essential features to eliminate trivial errors and increase programming efficiency. While these features are, to a degree, already available in Jupyter notebooks, the fact that our participants still requested them suggests that they are not prominent enough or hard to discover and use.

Likewise, the output, even from successful execution, was criticized for being hard to understand at times, but again, this is more of an issue of the specific implementation of the TensorFlow/Keras functions, which participants used, while more structured and natural language output is already possible and implemented by other languages and libraries.

\subsubsection{For Graphical Programming}
As it was only a prototypical implementation for the purpose of this study, naturally, the graphical programming tool also received its share of suggestions and feature requests. While this may be partly due to its lack of maturity, this can still show the priority of the participants.

Concerning feedback from the code execution, the feedback was that successful computations need not necessarily be displayed spatially close to the source code, like below Jupyter cells, but a single output terminal was adequate. For errors, however, the participants highlighted how an clear visual indicator at the erroneous location is desirable, to quickly identify the source of the error, while the output from the error message may not be immediately helpful.

Regarding the adoption of such a graphical tool, the consensus appears to be that participants were willing to give such a tool a try, but did not expect it to replace their familiar work environment. One reason they named for this, was that they expected no dramatic productivity benefit.
If participants already used graphical tools, they typically did so for external motivations, e.g. because it was  required by regulations or a legacy project required their use. For their everyday work, both for creating and using data-driven systems and in other scenarios, they were reluctant to use graphical programming tools. The justified this with the fact that they had a set of tools with which they are familiar, and adopting a new tool can be a time investment with uncertain revenue.
The participants did suggest two scenarios, though, where a graphical programming environment may find adoption: first for education or novices, as is a common idea described above. Second, they saw the high-level presentation as beneficial for setting up a data processing pipeline. In their expectation, they would, however, switch back to textual code for fine-tuning. Thus, they suggested some form of on-demand translation or switching between the different presentations to use different paradigms in different phases of the development project.

\section{Discussion}
This feedback, overall, paints a mixed picture. Both tools, Jupyter notebooks and our prototype of a graphical programming tool, appear to be adequate for the task at hand. Still, there are notable differences.

From an objective perspective, the significant differences in task completion time would suggest that direct manipulation is the superior way to create a data-driven application. With this paradigm, participants also asked fewer questions, which, to a degree, also suggests that they may have had less trouble, which can be a consequence of the more constrained nature of the graphical UI. And yet, the additional feedback is not as clear.

When judging the tools in isolation, the one category in which we observed a significant difference was the category of error prevention. While this is corroborated by fewer questions during use, it is an almost necessary consequence of the design: text input has absolute flexibility, while the graphical UI constrains the options. So, it is much easier to type an incorrect command accidentally. Likewise, when the developer has to remember all commands, they are more prone to make mistakes or not remember the correct parameters, values, etc.

However, the perception differed more strongly when we showed the alternative tool to the participants. While in the graphical tool was rated superior in appearance and how easy it is to learn, the text-based coding was rated to offer better flexibility.
Notably though, even in other categories, like consistency or how it supports exploration, the opinion of the participants shifted considerably, which suggests a strong anchoring bias, i.e., the perception of a tool is skewed by what the participants compare it to. Outside our study, this could mean that any new tool needs to measure up to what developers already use and with which they are very familiar. Therefore, it is not surprising that the adoption of alternative tools like graphical programming tools in general software development and data-driven development is sluggish at best and happens only in small incremental changes.

This is exacerbated by the opinion, as voiced by our participants, but also beyond that, that graphical programming should be targeted at novices or for learning, while the traditional text editor is the tool of choice for experts. Even if participants could identify the potential benefits of different forms of interaction and presentation, multiple suggested that any such benefit would not apply to them. In fact, they, for example, expect to be faster with written code, which the data in our study does not support. This perception may hold true for the pure input method, but this fails to consider contextual factors such as potentially difficult debugging, documentation, etc.

Of course, the feedback provided by the participants focuses on the two tools used in the study, which were deliberately limited in features. Additional support mechanisms like good auto-completion can make textual input faster, but the same potential for improvement also applies to other interaction paradigms beyond typing code.
While the support mechanisms for text are naturally more apparent, some of them may be adapted to other forms like graphical programming, or they could enable completely new support mechanisms. Certainly, this is not without challenges and as previously mentioned, graphical programming can become overwhelming with a growing number of features.
Therefore, striking a balance between offering enough support but remaining unobtrusive is important.

Generally though, the feedback from our study demonstrates that the lack of adoption is not a consequence of inherently inferior usability. The recorded metrics, in fact, show that a graphical programming environment with an equivalent level of features can be viewed on par and even superior in some aspects. Yet, even though our participants saw the potential benefits for productivity and effectiveness, they remain reluctant to use different types of tools, partially biased by their prior experience and habituation. If we hope to make the benefits of alternative forms of interaction more accessible, maintaining an adequate level of usability appears to be only a prerequisite. Since software developers seem to value some aspects of their user experience more, particularly flexibility and being in control of their software, this may be an aspect that needs to be addressed specifically.

A potentially simple way to facilitate this is to give developers more choice: if a tool supports different forms of interaction to create a data-driven application, e.g., via graphical and text-based view, developers can select which is most suitable in a given situation and switch between them. While the simple availability may convince some, it is still very possible that developers simply decide to fall back on what they know and do not utilize their options. The increased functionality also adds complexity, so, as previously described by \citet{PETRE95}, some expertise will be necessary to leverage this effectively.
It also remains yet unclear how such hybrid systems best consolidate the options to prevent this. More interactions and functionality easily runs the risk of adding complexity and confusion. There are many alternatives for combining these forms of interaction, be it side-by-side, as a mutually exclusive selection, or any form in between. Which of these or whether any of these have desirable effects should be explored in the future. Jupyter notebooks could be a viable platform for this in the future. While the core notebooks currently offer only hybrid output, extensions with more interactive widgets already exist (e.g., \cite{kery2020mage}) and adding to this to allow for fully hybrid interaction certainly seems viable.

Given how used many developers already are to their text editors or IDEs, for wide adoption, other forms of interaction in software development must be advertised as a viable alternative for productive, professional use.
Combining this with the opinion that graphical programming may be suitable for education, early Computer Science education may be a point where such a change in perception can be fostered. If a new generation of developers is introduced to not just the traditional text editor or Integrated Development Environments (IDEs), but also consistently uses other tools in productive use, they may become normalized and be just one tool in a developer's toolbox.

Usage for real-world tasks with long-term observations could then yield insights into how a wider tool choice affects work practices over time and how this, in turn, can affect tool design. While this is possible in the few domains who already use graphical tools, observing this in professional, large-scale data-driven development is, unfortunately, currently not yet possible for lack of adoption. This area, though, could provide some interesting insights, particularly where deep neural networks are used. They are an especially interesting case for other forms of programming and presentation, since in them a lot of information and effort is highly condensed in an often opaque fashion, making them hard to grasp. Since people thus already often communicate about them with visualizations, a graphical presentation could make them more transparent and explicit. Still, their size leads to different challenges of presentation. Feasibility considerations limited our study to use the MNIST example, which also allowed for fairly controlled observations. While it contains all fundamental steps of the typical workflow, this constrained example may not elicit all the nuances of data-driven development, though. With larger systems like deep neural networks and in field studies, it will be more challenging to compare equivalent systems implemented using different tools as we did in our study, but the feedback could still help refine our understanding where the discrepancy in perception of such tools originates and which features make graphical programming viable to handle the complexity of large neural networks.

\section{Conclusion}

This paper contributes insights into the design of software development tools for data-driven development. Our experiments compare textual code-based programming to  graphical programming as an alternative rich form of interaction. They show the latter's benefits, for example, faster task completion and better error prevention. In addition, the participants in our study can identify further benefits, especially related to learning and the ability to grasp the general structure of a complex data-driven application quickly. In isolation, the perception of this form of interaction is not much different from that of typing out code, though, but when presented with the alternative, this difference becomes more pronounced though, indicating an anchoring bias.

At times, participants' subjective perception does not reflect the actual situation, and participants dismiss some potential benefits but dismiss them as only applicable to others. All this, coupled with an attitude to ``never change a running system'', after all, current methods are capable of producing working software, could be an indicator why not just in data-driven development but in software development in general, a large number of tools with rich interactions are acknowledged but play no significant role in productive use. Our study also highlights how general usability appears to play a minor role, while certain aspects like flexibility seem more important.

One attempt to overcome this from the perspective of interaction design may be to lean into the aspect of flexibility and developers to choose from different forms of interaction, both traditional and novel, depending on the usage context and current developer's needs. How exactly code and other forms of rich interaction, like direct manipulation in our prototype, can be combined not just as replacement for each other but in synergy remains an open question. Since Jupyter, as the popular choice for this kind of programming, was still perceived relatively positively and allows for some extensibility, it may be a good starting point. This can also mitigate any overhead which a completely new tool could bring. It remains to be seen, though, how far we can take the existing tools, which were designed for text-based input, and whether, at some point, the ever-growing complexity of data-driven application becomes sufficient motivation to adopt alternative tool paradigms.

\bibliographystyle{ACM-Reference-Format}
\bibliography{references}

\end{document}